\documentclass[a4paper,fleqn]{cas-dc}

\usepackage[numbers]{natbib}

\def\tsc#1{\csdef{#1}{\textsc{\lowercase{#1}}\xspace}}
\tsc{WGM}
\tsc{QE}
\tsc{EP}
\tsc{PMS}
\tsc{BEC}
\tsc{DE}

\def\be{\begin{equation}}
	\def\ee{\end{equation}}
\newcommand{\bel}[1]{\begin{eqnarray}\label{#1}}
	\newcommand{\eel}{\end{eqnarray}}
\def\barr{\begin{array}}
	\def\earr{\end{array}}
\def\beq{\begin{eqnarray}}
	\def\eeq{\end{eqnarray}}
\def\bfig{\begin{figure}}
	\def\efig{\end{figure}}

\newcommand{\nn}{\nonumber}
\newcommand{\f}[2]{\frac{#1}{#2}}
\newcommand{\onehalf}{{\nicefrac{1}{2}}}

\newcommand{\p}{\partial}

\newcommand{\rf}[1]{Eq.~(\ref{#1})}

\def\g{\gamma}

\def\c{\chi}
 
\usepackage{amssymb}

\usepackage{graphicx}
\usepackage{amsmath}
\usepackage{subfigure}
\usepackage{bbold}
\usepackage{yfonts}
\usepackage{placeins}
\usepackage{bm} 
\usepackage{nicefrac}
\usepackage{slashed}

\newcommand{\bea}{\begin{eqnarray}}
\newcommand{\eea}{\end{eqnarray}}

\newcommand{\EQ}[1]{Eq.~(\ref{#1})}
\newcommand{\EQn}[1]{(\ref{#1})}

\newcommand{\EQSM}[2]{Eqs.~(\ref{#1})--(\ref{#2})}

\newcommand{\CIT}[1]{Ref.~\cite{#1}} 
 
\newcommand{\CITn}[1]{\cite{#1}}

\newcommand{\bv}{{\boldsymbol b}} 
\newcommand{\ev}{{\boldsymbol e}} 
\newcommand{\nv}{{\boldsymbol n}}

\newcommand{\vv}{{\boldsymbol v}}
\newcommand{\pv}{{\boldsymbol p}}

\newcommand{\Pv}{{\boldsymbol P}}

\newcommand{\gammav}{{\boldsymbol \gamma}}

\newcommand\sigv{{\boldsymbol \sigma}}
\newcommand{\zetav}{{\boldsymbol \zeta}}

\def\kv{{\boldsymbol {k}}}

\newcommand{\trt}{{\rm tr_2}}
\newcommand{\trf}{{\rm tr_4}}

\def\omnL{\omega_{\mu\nu}}
\def\omnU{\omega^{\mu\nu}}

\def\omnUD{\tilde {\omega}^{\mu\nu}}

\def\SmnU{{\Sigma}^{\mu\nu}}
\def\SmnL{{\Sigma}_{\mu\nu}}
 
\def\S0iU{{\Sigma}^{0i}}

\def\n0{n_{(0)}}
\def\e0{\varepsilon_{(0)}}
\def\P0{P_{(0)}}

\def\fplusrs{f^+_{rs}}

\def\fminusrs{f^-_{rs}}

\def\ubarrp{{\bar u}_r(p)}
\def\ubarsp{{\bar u}_s(p)}
\def\usp{u_s(p)}
\def\urp{u_r(p)}

\def\vbarrp{{\bar v}_r(p)}
\def\vbarsp{{\bar v}_s(p)}
\def\vsp{v_s(p)}
\def\vrp{v_r(p)}

\def\Wpxk{{\cal W}^{+}(x,k)}
\def\Wmxk{{\cal W}^{-}(x,k)}
\def\Wpmxk{{\cal W}^{\pm}(x,k)}
\def\Wxk{{\cal W}(x,k)}
\def\Weqxk{{\cal W}_{\rm eq}(x,k)}

\def\Fpmxk{{\cal F}^{\pm}(x,k)}

\def\Feqpmxk{{\cal F}^{\pm}_{\rm eq}(x,k)}

\def\Ppmxk{{\cal P}^{\pm}(x,k)}

\begin{document}
\let\WriteBookmarks\relax
\def\floatpagepagefraction{1}
\def\textpagefraction{.001}
\shorttitle{Spin chemical potential for relativistic particles with spin 1/2}
\shortauthors{W. Florkowski et~al.}

\title [mode = title]{Spin chemical potential for relativistic particles with spin $\onehalf$}                      



\author[1]{Wojciech Florkowski}[
                       orcid=0000-0002-9215-0238
                        ]
\cormark[1]
\ead{wojciech.florkowski@uj.edu.pl}


\address[1]{M. Smoluchowski Institute of Physics, Jagiellonian University, PL-30-348 Krakow, Poland}

\author[2]{Avdhesh Kumar}[orcid=0000-0003-1335-8880]

\ead{avdhesh.kumar@ifj.edu.pl}

\author[2]{Radoslaw Ryblewski}[ orcid=0000-0003-3094-7863
                        %
   ]
\ead{radoslaw.ryblewski@ifj.edu.pl}


\address[2]{Institute of Nuclear Physics Polish Academy of Sciences, PL-31-342 Krakow, Poland}



\cortext[cor1]{Corresponding author}


\begin{abstract}
We analyze algebraic structure of a relativistic semi-classical Wigner function of particles with spin~$\onehalf$ and show that it consistently includes information about the spin density matrix both in two-dimensional spin and four-dimensional spinor spaces. This result is subsequently used to explore various forms of equilibrium functions that differ by specific incorporation of spin chemical potential. We argue that a scalar spin chemical potential should be momentum dependent, while its tensor form may be a function of space-time coordinates only. This allows for the use of the tensor form in local thermodynamic relations. We furthermore show how scalar and tensor forms can be linked to each other. 
\end{abstract}



\begin{keywords}
Wigner function \sep spin density matrix \sep spin polarization tensor \sep spin chemical potential \sep relativistic heavy-ion collisions
\end{keywords}

\maketitle
\section{Introduction}
In this letter we first analyze algebraic structure of a relativistic semi-classical Wigner function for massive particles with spin~$\onehalf$. We restrict our considerations to the leading order of expansion in $\hbar$ and show that it consistently includes information about the spin density matrix both in two dimensional spin and four dimensional spinor spaces. This consistency strongly relies on the fact that the two-by-two spin density matrix operates with quantities defined in the particle rest frame. 

In the next step, we study equilibrium Wigner functions that differ by the form of spin chemical potential. We demonstrate that a scalar spin chemical potential frequently used in the literature \CITn{Fang:2016vpj,Weickgenannt:2019dks,Schaefer:2016yzd} should be momentum dependent. In contrast, a tensor spin chemical potential, introduced in new studies of hydrodynamics with spin~\CITn{Florkowski:2017ruc}, may be a function of space-time coordinates only. This allows for the use of the tensor form in local thermodynamic relations. We furthermore show how scalar and tensor forms can be linked to each other, provided the polarization effects are small. 

Our results show that the introduction of a scalar spin chemical potential is quite arbitrary. In contrast, the tensor form has much better physical motivation, as it plays role of the Lagrange multiplier(s) coupled to  angular momentum, which is a conserved quantity~\CITn{Becattini:2018duy}. In this context, the name ``chemical potential'' seems to be better justified, as we may interpret the total angular momentum as a ``conserved charge''.

We expect that our results will be helpful for better understanding of equilibrium properties of particles with spin. This is important for development of hydrodynamic and kinetic theories for such systems, and  is very much desirable in the context of the spin polarization measurements in heavy-ion collisions~\CITn{STAR:2017ckg}. The latter revealed a non-zero effect for $\Lambda$ hyperons, with a momentum dependence of polarization still waiting for a convincing theoretical explanation \cite{Niida:2018hfw}. 

Our paper also clarifies the Lorentz structure of different quantities describing spin densities, therefore, it may be useful for future studies dealing with the relativistic spin dynamics. Although some of the formulas presented below were obtained before, to our knowledge, no attempt has been made before to directly link them all and explain their physical interpretation. 

In Secs. 2--4 we analyze the structure of the Wigner functions not referring to any concept of equilibrium. Only in Secs. 5--9, which are central for our work, we discuss various equilibrium forms. We conclude in Sec.~10.

We use the metric tensor with the signature $(+,-,-,-)$ and the Levi-Civita symbol with $\epsilon^{0123}=+1$. The trace over spinor (spin) indices is denoted by $\trf$ ($\trt$). The conventions regarding the spinors and several useful relations are collected in Appendix~\ref{sec:usefulid}.

\section{Semi-classical Wigner functions}

Our starting point are Wigner functions for particles and antiparticles, ${\cal W}^\pm(x,k)$, obtained in the leading order of the semi-classical expansion \CITn{DeGroot:1980dk},
\bea
\Wpxk = \frac{1}{2} \!\sum_{r,s=1}^2 \int \!\!\! dP\,
\delta^{(4)}(k\!-\!p) u^r(p) {\bar u}^s(p) f^+_{rs}(x,p),\!\!\!
\label{eq:Weqpxk}
\eea
\bea
\Wmxk = -\frac{1}{2} \!\sum_{r,s=1}^2 \int \!\!\!dP\,
\delta^{(4)}(k\!+\!p) v^s(p) {\bar v}^r(p) f^-_{rs}(x,p).\!\!\!
\label{eq:Weqmxk}
\eea
Here $m$ is the (anti)particle mass, $k$ is the four-momentum, and $dP$ is the Lorentz invariant integration measure $dP = d^3p/((2 \pi )^3 E_p)$, where $E_p = \sqrt{m^2 + \pv^2}$ is the on-mass-shell energy and $p^\mu = (E_p, \pv)$.  The objects $\urp$ and $\vrp$ are Dirac bispinors with the spin indices $r$ and $s$ running from 1~to~2 and the normalizations: $\ubarrp \usp=\,2m\, \delta_{rs}$ and $\vbarrp \vsp=-\,2m\, \delta_{rs}$. We note that a minus sign and a different ordering of spin indices are used in \rf{eq:Weqmxk} compared to \rf{eq:Weqpxk}. 

The total Wigner function becomes a sum of the particle and antiparticle contributions $\Wxk = \Wpxk + \Wmxk$. One can easily check that $(\slashed{k} -m)\Weqxk = 0$, as required for the leading-order term of the Wigner function in the $\hbar$ expansion~\cite{Vasak:1987um,PhysRevD.46.4603,Zhuang:1995pd,Florkowski:1995ei}. 

The functions $\Wpxk$ and $\Wmxk$ can be expressed with the help of 16~independent generators of the Clifford algebra~\cite{Vasak:1987um,Itzykson:1980rh},
\beq
\Wpmxk \!\!\!&=&\!\!\! \f{1}{4} \left[ \Fpmxk + i \gamma_5 \Ppmxk + \gamma^\mu {\cal V}^\pm_{\mu}(x,k) \right. \nn \\
&& \left.  + \gamma_5 \gamma^\mu {\cal A}^\pm_{\mu}(x,k)
+ \SmnU {\cal S}^\pm_{ \mu \nu}(x,k) \right].
\label{eq:wig_expansion}
\eeq
Here $\SmnU$ is the Dirac spin operator, $\SmnU = \f{i}{4} [\gamma^\mu,\gamma^\nu]$. In the leading order of semi-classical expansion, one can check that only scalar and axial-vector coefficient functions are independent. The other coefficients are expressed in terms of ${\cal F}^\pm = \trf \left[\Wpmxk\right]$ and $ {\cal A}^\pm_{\mu}= \trf \left[\gamma_{\mu}\gamma^5 \Wpmxk\right]$ by the following expressions~\cite{Vasak:1987um}:
\beq
{\cal P}^\pm(x,k) \!\!\!&=&\!\!\! -i\,\trf\left[\gamma^5\Wpmxk\right] = 0,
\label{eq:Peq} \\
{\cal V}^\pm_{\mu}(x,k) \!\!\!&=&\!\!\! \trf\left[\gamma_{\mu}\Wpmxk\right] = \frac{k_\mu}{m} {\cal F}^\pm(x,k),
\label{eq:rVeq} \\
{\cal S}^\pm_{\mu \nu}(x,k) \!\!\!&=&\!\!\! 2\,\trf\left[\SmnL \Wpmxk\right] = -\frac{1}{m} \epsilon_{\mu\nu \alpha \beta} k^\alpha {\cal A}^{\pm\beta}(x,k).\nn\\
\label{eq:rSeq}
\eeq
This set of equations should be supplemented by a subsidiary condition $k_\beta {\cal A}^{\pm\, \beta}(x,k) = 0$.

Using \EQSM{eq:Peq}{eq:rSeq} in the definition of the Wigner function one finds
\beq
\Wpmxk &=& \frac{1}{4m} (m+\slashed{k}) \left[{\cal F}^\pm + \gamma_5 \gamma_\beta {\cal A}^{\pm\, \beta} \right].
\label{eq:vasak1}
\eeq 
In this way we reproduce Eq.~(5.44) from \CIT{Vasak:1987um} (note a different normalization and an opposite sign in front of ${\cal A}^{\pm\, \beta}$, which is a consequence of different conventions used in~\CITn{Vasak:1987um}, see also \CITn{PhysRevD.46.4603}).


\section{Spin density matrix}
The functions $f^+_{rs}(x,p)$ and $f^-_{rs}(x,p)$ play role of the spin density matrices. They are  two-by-two Hermitian matrices which can be generally decomposed as~\CITn{Florkowski:2017dyn}
\bel{fpm}
f_{rs}^\pm(x,p) = f^\pm_0(x,p) \left[\delta_{rs}  + \zetav_\ast^\pm(x,\pv) \cdot \sigv_{rs} \right].
\eel
Here $\sigv$ denotes a three-vector consisting of three Pauli matrices. The three-vector $\zetav_\ast^\pm(x,\pv)$ can be interpreted as a spatial part of the polarization four-vector $\zeta^{\pm \mu}_\ast(x,p)$, with a vanishing zeroth component,~\footnote{We follow here the arguments discussed in \CITn{Florkowski:2017dyn}, where in the case of small polarization the identification $\Pv = -2 \zetav^{\pm}_\ast$ should be made.} 
\bel{eq:Pmu}
\zeta^{\pm \mu}_\ast(x,p)=\left(0,\zetav_\ast^\pm(x,\pv)\right).
\eel
The average polarization vector is defined by the formula
\bea
\left\langle \zetav_\ast^\pm(x,\pv) \right\rangle =  \f{1}{2} \f{ \trt \left( f^\pm \sigv\right)  }{\trt \left( f^\pm \right) }  = 
\frac{1}{2} \zetav_\ast^\pm(x,\pv) .
\label{eq:avPv1}
\eea
Several important points should be emphasized here: 
\begin{itemize}
    \item The polarization three-vector $\zetav^\pm_\ast$ describes spin polarization in the particle (antiparticle) rest frame (PRF), i.e., in the frame where $p^\mu = (m,0,0,0)$. We denote this frame by asterisk~\CITn{Leader:2001gr}. 
    \item The measurements of the spin polarization vary between $-1/2$ and $+1/2$, hence, $| \zetav^\pm_\ast|~\leq~1$. The particle spin states with $| \zetav_\ast^\pm| = 1$ correspond to pure states, while the cases with $| \zetav_\ast^\pm| < 1$ desribe mixed states. 
    \item The functions $f^\pm_0$ contain information averaged over spin degrees of freedom. Hence, it is tempting to write them as sums of the distributions of particles with spin up and down. We thoroughly discuss this point below.
    \item We stress that $\zetav^\pm_\ast$ is a function of space-time coordinates and three-momentum of particles, $\zetav^\pm_\ast = \zetav^\pm_\ast(x,\pv)$. The quantity $\zetav^\pm_\ast(x,\pv)$, after averaging over the space-time region where particles are produced, becomes a directly measured observable. This happens, for example, in the case of $\Lambda$ spin polarization measured in heavy-ion collisions. 
\end{itemize}
To transform the PRF components of any four-vector to the laboratory (LAB) frame, we use the so called canonical boost $\Lambda^\mu_{\,\,\,\nu}(\vv_p)$ (see, for example, Eq.~(45) in \CIT{Florkowski:2017dyn}).  In the case of the four-vector $\zeta^\mu_\ast(x,\pv)$, this leads to the formula
\bel{eq:Zetamu}
\zeta^\mu_\pm = \Lambda^\mu_{\,\,\,\nu}(\vv_p) \zeta^\nu_{\pm \ast} = \left(\frac{\pv \cdot \zetav^\pm_\ast}{m} , \zetav^\pm_\ast + \frac{\pv \cdot \zetav^\pm_\ast}{m (E_p + m)} \pv \right).
\eel

In relativistic quantum mechanics and quantum field theory, one deals with the spin densities defined in the spinor space. It is interesting to show that the expression \EQn{eq:vasak1} is proportional to such densities. With the explicit forms of matrix elements given in Appendix \ref{sec:usefulid} we find
\beq 
\hspace{-0.8cm}{\cal F}^\pm (x,k) \!\!\!&=&\!\!\! 2 m \int dP \delta^{(4)}(k \mp p) f^\pm_0(x,p), 
\label{eq:Fv} \\
\hspace{-0.8cm}{\cal A}^{\pm \beta}(x,k) \!\!\!&=&\!\!\! 2 m \int dP \delta^{(4)}(k \mp p) f^\pm_0(x,p) \zeta^{\pm \beta}(x,\pv),
\label{eq:Av}
\eeq 
and
\beq
\hspace{-0.8cm}\Wpmxk &=& \pm \int dP \delta^{(4)}(k \mp p) f^\pm_0(x,p) \rho^\pm(x,\pv).
\label{eq:LL1}
\eeq 
Here we have introduced the four-dimensional matrices
\beq 
\rho^\pm(x,\pv) = \frac{1}{2} (\slashed{p} \pm m ) \left(1 +  \gamma_5 \slashed{\zeta}^\pm \right),
\label{eq:rhopm}
\eeq 
which exactly agree with the definitions of the polarization spin matrices given in~\CITn{Berestetsky:1982aq}.~\footnote{Note that the convention for $\gamma_5$ used in~\CITn{Berestetsky:1982aq} differs by sign from ours, see Appendix~\ref{sec:usefulid}. Note also that our results are obtained by a straightforward calculation of the matrix elements rather than by a diagonalization of the matrix $f^\pm_{rs}$, what has been done in~\CIT{Weickgenannt:2019dks}.} 

\section{Scalar and axial components}
 Doing the integral over three-momentum in \EQ{eq:Fv} one finds
\beq 
{\cal F}(x,k) = \frac{4m}{(2\pi)^3} \delta(k^2-m^2)
F(x,k)
\label{eq:Fv1}
\eeq 
with
\beq 
F(x,k) = 
\left[ \theta(k^0) f^+_0(x,k) + \theta(-k^0) f^-_0(x,-k) 
\right].
\label{eq:Fxk}
\eeq 
Similar decomposition can be obtained for the axial component, however, in this case it is useful to introduce yet another form of the polarization vectors $\zeta^{\pm \beta}$. Since they are space-like, we can write them in the form~\footnote{We note that the $\pm$ signs in the definition (\ref{eq:zetapm}) are conventional and the minus sign in (\ref{eq:zetapm}) compensates the minus sign in the middle of the right-hand side of (\ref{eq:ns}).}
\beq 
\zeta^{\pm \beta}(x,\pv) = \pm \zeta^\pm(x,\pv) n^{\pm \beta}(x,\pv), 
\label{eq:zetapm}
\eeq
where $n^{\pm \beta}(x,\pv) n^{\pm}_\beta(x,\pv) = -1$ and
\beq 
\zeta^\pm(x,\pv) = \sqrt{-\zeta^{\pm \beta}(x,\pv)  \zeta^{\pm}_{\beta}(x,\pv)} = |\zetav_\ast^\pm|.
\label{eq:mus}
\eeq 
Here we used the fact that the scalar product can be calculated in any frame and chose PRF. The explicit form of $n^\mu_\pm $ is
\bel{eq:nmu}
n^\mu_\pm(x,\pv) = \pm \left(\frac{\pv \cdot \nv^\pm_\ast}{m} , \nv^\pm_\ast + \frac{\pv \cdot \nv^\pm_\ast}{m (E_p + m)} \pv \right),
\eel
where
\beq 
\nv^\pm_\ast(x,\pv) = \frac{\zetav^\pm_\ast(x,\pv)}{|\zetav^\pm_\ast(x,\pv)|} 
= \frac{\zetav^\pm_\ast(x,\pv)}{\zeta^\pm(x,\pv)}.
\eeq 
We observe that the three-vectors $\nv^\pm_\ast(x,\pv)$ describe the direction of mean polarization of particles with momentum $\pv$ (measured in PRF), while the positive quantity $\zeta^\pm(x,\pv)$ defines the magnitude of spin polarization. 

We stress again that the case $\zeta^\pm(x,\pv) = 1$ corresponds to a pure state, while the case $\zeta^\pm(x,\pv) < 1$ describes a mixed state. Thus, in most of the cases, the three-vector $\nv^\pm_\ast(x,\pv)$ cannot be interpreted as an arbitrary quantization axis. It describes the {\it mean direction} obtained by measurements of spin projections of many particles along three independent directions. 

Performing the integral over three-momentum in \EQ{eq:Av} and using the notation introduced above one gets
\beq 
{\cal A}^{\beta}(x,k) = \frac{4m}{(2\pi)^3}  \delta(k^2-m^2) n^\beta(x,k) A(x,k)
\label{eq:Av2}
\eeq 
where
\beq 
n^\beta = \theta(k^0) n^{+ \beta}(x,\kv) - \theta(-k^0) n^{- \beta}(x,-\kv)
\label{eq:ns}
\eeq 
and
\beq 
A(x,k)&=&\left[\theta(k^0) f^+_0(x,k) \zeta^+(x,\kv)\right.\nn\\&&+\left.\theta(-k^0) f^-_0(x,-k) \zeta^-(x,-\kv) \right].
\label{eq:Axk}
\eeq

At this point it is useful to compare our framework with previous, similar studies. In particular, one can check that \EQ{eq:ns} is consistent with the expressions (26) and (27) obtained in \CIT{Weickgenannt:2019dks}, provided the vectors $\nv^\pm_\ast(x,\kv)$ are identified with the vectors $\nv^\pm$ defined therein. A~subtle difference exists, however, since the vectors $\nv^\pm_\ast(x,\kv)$ do depend on $\kv$, hence, the vectors $\nv^\pm$ in \CIT{Weickgenannt:2019dks} should be also consistently treated as functions of $\kv$.

Similar comments apply to \CIT{Fang:2016vpj}. Our results agree with Eqs.~(26) and (28) in \CIT{Fang:2016vpj}, if the vector $\nv$ defined there is simultaneously equal to  $\nv^+_\ast(x,\kv)$ and $\nv^-_\ast(x,-\kv)$. Thus, we agree with \CIT{Fang:2016vpj} only if $\nv^+_\ast(x,\kv) = \nv^-_\ast(x,-\kv)$. The last condition represents a constraint on the most likely directions of polarization vectors for particles and antiparticles. We come back to its interpretation below \EQ{Aveb}.

Besides the two vectors $n^{\pm \beta}$, the system under consideration is described by the four scalar functions: $f^\pm_0$ and $\zeta^\pm$. They can be conveniently reorganized to describe particles with spins up and down along the direction set by unit vectors $n^{\pm \beta}$. This can be done with the help of the definition
\beq 
f^\pm_{0s}(x, \pm k) = \frac{1}{2} f^\pm_0(x,\pm k)\left(1 + s \zeta^\pm(x,\pm \kv)\right),
\label{eq:spinfun}
\eeq 
where $s=\pm 1$ denotes the spin direction. Note that we have $0 \leq \zeta^\pm(x,\pm \kv) \leq 1$, hence $f^\pm_{0s}(x, \pm k)$ is positive if $f^\pm_0(x,\pm k) > 0$. Equation \EQn{eq:spinfun} allows us to rewrite Eqs.~(\ref{eq:Fxk}) and (\ref{eq:Axk}) as
\beq 
F(x,k) &=& 
\left[ \theta(k^0) \left(f^+_{0+}(x,k)+f^+_{0-}(x,k)\right) \right. \label{eq:Fxk1} \\
&& \left. + \theta(-k^0) \left(\vphantom{f^+_{0+}}f^-_{0+}(x,-k)+f^-_{0-}(x,-k)\right) 
\right] \nn
\eeq 
and 
\beq 
A(x,k) &=& 
\left[ \theta(k^0) \left(f^+_{0+}(x,k)-f^+_{0-}(x,k)\right) \right. \label{eq:Axk1} \\
&& \left. + \theta(-k^0) \left(\vphantom{f^+_{0+}}f^-_{0+}(x,-k)-f^-_{0-}(x,-k)\right) 
\right]. \nn
\eeq 

\section{Equilibrium Wigner functions}
So far we have not addressed the fact that our Wigner function describes a system of particles with spin in equilibrium. As a matter of fact, different forms of such functions are proposed in the literature and one of the main aims of this work is to examine them and check their internal consistency connected with relativistic covariance and physical interpretation of the spin polarization measurements. 

The optimal situation would be to derive an equilibrium form from the considerations that analyze either entropy production or the form of collision terms for particles with spin. As such calculations are not available at the moment, various  discussions of the equilibrium for particles with spin have to make use of different arguments, usually combined together, to conclude about the acceptable forms of the equilibrium functions. These functions necessarily invoke certain forms of the spin chemical potential, hence, the issue of choosing the correct equilibrium form is connected with the introduction of the appropriate spin chemical potential.   

Some support in this respect comes from the analysis of kinetic theory with classical description of spin. With the arguments about the locality of the classical collision term, one can construct for this case an equilibrium distribution function that naturally involves a tensor spin chemical potential~\CITn{Florkowski:2018fap}. We come back to this point below and turn to a discussion of specific equilibrium Wigner functions now.

\section{Scalar spin chemical potential}
Since $f_0^\pm$ describes an average over the spin components, see \EQ{eq:Fxk1}, it seems natural to assume that $f_{0s}^\pm$ has the form of the standard equilibrium function depending on the flow vector $u^\mu$, temperature $T$, chemical potential $\mu_e$ connected with the conservation of charge, and an additional spin chemical potential~$\mu^\pm$ that controls the relative number of particles (plus sign) or antiparticles (minus sign) with spin up and down, namely 
\beq 
f^\pm_{0s} \!\!\! &=&\!\!\!\!\!
\frac{1}{2} \left[\exp\left(\frac{p \cdot u \mp \mu_e - s \mu^\pm}{T}  \right)\!+\!1 \right]^{-1} \nn \\
\!\!\! &\approx&\!\!\!\!\!
\frac{1}{2}\left[\exp\left(\frac{p \cdot u \mp \mu_e }{T}  \right)\!+\!1 \right]^{-1}
\left(1\!+\!\frac{s \, \frac{\mu^\pm}{T}}{
1 + \exp\left(\frac{- p \cdot u \pm \mu_e }{T}  \right) } \right) \nn \\
\!\!\! &\approx&\!\!\!\!\!
\frac{1}{2}\exp\left(- \frac{p \cdot u \pm \mu_e }{T}  \right) \left( 1 + s \, \frac{\mu^\pm}{T} \right). \label{eq:scalarspinchempot}
\label{eq:FD}
\eeq 
Since we are dealing with spin-$\onehalf$ particles, in the first line of \EQ{eq:FD} we have used the Fermi-Dirac statistics. Clearly, this form does not comply with the general structure given by \EQn{eq:spinfun}. Matching between \EQ{eq:FD} and \EQn{eq:spinfun} can be reached, however, if the effects of polarization are small, namely, for $\mu^\pm/T \ll 1$, which yields the second line of \EQ{eq:FD}. Eventually, the relation between $\mu^\pm$ and $\zeta^\pm(x,\pv)$ becomes quite simple for the case of the Boltzmann statistics, shown in the third line of \EQ{eq:FD}. In this case we find
\beq 
\mu^\pm = T(x) \zeta^\pm(x,\pv).
\label{eq:match}
\eeq 

The first lesson we can take from the above discussion is that inclusion of the scalar spin chemical potential in the standard equilibrium distribution functions makes sense only if the effects of spin polarization are small, which is quantified by the condition $\mu^\pm/T \ll 1$. More importantly, $\mu^\pm$ should be treated as a function of momentum of particles with spin, $\mu^\pm = \mu^\pm(x,\pv)$ . Consequently, it cannot be used in a traditional way in thermodynamic identities.

The origin of this difficulty is a simple fact that the spin polarization of relativistic massive particles is always defined in their rest frames, hence, different boosts should be applied to particles with different three-momenta in order to determine their spin polarization. This dependence is reflected in the momentum dependence of $\mu^\pm$, which eventually makes it a badly defined quantity from the thermodynamic point of view. Clearly, the problems outlined above disappear in the non-relativistic limit. 

\section{Tensor spin chemical potential}

In~\CIT{Becattini:2013fla}, the following local equilibrium Wigner functions were introduced, 
\beq
\fplusrs \!\!\!\!&=&\!\!\!\!\!
\frac{1}{2m} \ubarrp \exp\left[- p \cdot \beta + \xi_e + \f{1}{2} \varpi_{\mu\nu}  \SmnU \right]  \usp, \nn  \\
\fminusrs \!\!\!\!&=&\!\!\!\!\!
- \frac{1}{2m}\vbarsp \exp\left[-p \cdot \beta - \xi_e - \frac{1}{2} \varpi_{\mu\nu}  \SmnU \right]  \vrp, \nn \\ \label{eq:fbyFB}
\eeq
where $\xi_e = \mu_e/T$ and $\varpi_{\mu\nu}$ is thermal vorticity defined by the expression  $\varpi_{\mu\nu} = -(1/2) \left( \p_\mu \beta_\nu - \p_\nu \beta_\mu \right)$ with $\beta^\mu=u^\mu/T$. 

The equilibrium forms \EQn{eq:fbyFB} were subsequently used in \CIT{Florkowski:2017ruc} to construct relativistic hydrodynamics of particles with spin $\onehalf$. The main idea of \CIT{Florkowski:2017ruc} was to replace thermal vorticity in \EQ{eq:fbyFB} by the spin polarization tensor $\omega_{\mu\nu}$, whose dynamics should be determined by the conservation of angular momentum (instead of being tightly connected with thermal vorticity). The spin polarization tensor can be identified with the ratio $\Omega_{\mu\nu}/T$, where $\Omega_{\mu\nu}$ plays a role of a tensor spin chemical potential (both $\omega_{\mu\nu}$ and $\Omega_{\mu\nu}$ are rank two antisymmetric tensors that depend only on space and time coordinates, for brevity of notation we dominantly use $\omega_{\mu\nu}$ instead of $\Omega_{\mu\nu}$). 

If the components of $\omega_{\mu\nu}$ are small, the form of equilibrium Wigner function advocated in \CIT{Florkowski:2017ruc} agrees with \EQ{eq:vasak1} where one should use~\CITn{Florkowski:2018ahw}
\beq 
\Feqpmxk \!\!\!&=&\!\!\! 2 m \,\int dP\, \,e^{-\beta \cdot p \pm \xi_e}\,\,\delta^{(4)} (k\mp p)  \label{eq:FEeqpm}
\eeq 
and
\beq 
{\cal A}^\pm_{{\rm eq}, \mu}(x,k) \!\!\!&=&\!\!\! -\,\int dP\,e^{-\beta \cdot p \pm \xi_e}\,\,\delta^{(4)}(k\mp p)\, \tilde{\omega }_{\mu \nu}\,p^{\nu}.\nn\\
\label{eq:AEeqpm} 
\eeq
Here  ${\tilde \omega}_{\mu\nu}$ is the dual spin polarization tensor defined as ${\tilde \omega}_{\mu\nu} = (1/2) \epsilon_{\mu\nu\alpha\beta} \omega^{\alpha\beta}$.

One can notice that the approach proposed in ~\CIT{Becattini:2013fla} and extended in \CIT{Florkowski:2017ruc} introduces the same spin polarization tensor for particles and antiparticles. This makes sense if they are all in common equilibrium. As a matter of fact, the experimental data in heavy-ion collisions suggest that the spin polarization of $\Lambda$'s is the same as that of ${\bar \Lambda}$'s, which is interpreted as a consequence of a local thermodynamic equilibrium in which all particles take part. In the following we shall use the same tensor $\omega_{\mu\nu}$ for particles and antiparticles, although at this stage it is possible to introduce $\omega^+_{\mu\nu}$ and $\omega^-_{\mu\nu}$ which differ from each other.

The antisymmetric spin polarization tensor $\omnL$ can be always defined  in terms of electric- and magnetic-like three-vectors in LAB frame, $\ev = (e^1,e^2,e^3)$ and $\bv = (b^1,b^2,b^3)$. In this case, following the electrodynamic sign conventions of \CITn{Jackson:1998nia}, we write~\CITn{Florkowski:2017dyn}
\bel{omeb}
\omnL= 
\begin{bmatrix}
	0       &  e^1 & e^2 & e^3 \\
	-e^1  &  0    & -b^3 & b^2 \\
	-e^2  &  b^3 & 0 & -b^1 \\
	-e^3  & -b^2 & b^1 & 0
\end{bmatrix}.
\label{eq:omegaeb}
\eel
The dual spin polarization tensor is obtained from the components of $\omega_{\mu\nu}$ by replacements $\ev \to \bv$ and $\bv \to -\ev$. In \CIT{Florkowski:2017dyn} it was demonstrated that 
\beq 
\zetav^\pm_\ast(x,\pv) \!\!\!&=&\!\!\! -\f{1}{2 m} \left[  E_p \, \bv - \pv \times \ev - \f{\pv \cdot \bv}{E_p + m} \pv \right].
\label{Aveb}
\eeq 
Equation \EQn{Aveb} shows that the spin polarization vectors of particles and antiparticles are indeed the same (in equilibrium described with the help of the tensor chemical potential). This makes sense if they are in common equilibrium state. We have seen above that the condition $\nv^+_\ast(x,\kv) = \nv^-_\ast(x,-\kv)$ is used in \CIT{Fang:2016vpj}. For the tensor chemical potential this implies that $\ev =0$ in this case. The physical interpretation of this equation remains to be clarified. At the moment, we may notice that $\ev = 0$ in the global equilibrium states with a rigid rotation~\CITn{Becattini:2009wh}. 

Using \EQ{Aveb} in \EQ{eq:Zetamu} or by making a direct comparison of Eqs.~\EQn{eq:Av} and \EQn{eq:AEeqpm} we find the identification
\beq 
\zeta^\pm_\mu(x,\pv) = -\frac{1}{2m} {\tilde \omega}_{\mu\nu}(x)p^\nu.
\eeq 
In \CIT{Florkowski:2017dyn} it was also shown that the right-hand side of \EQ{Aveb} coincides with the value of the $\bv$ field determined in PRF, namely
\beq 
\zetav^\pm_\ast(x,\pv) = -\frac{1}{2} \bv_\ast(x,\pv).
\eeq 
This is a suggestive result indicating that for the spin polarization only the magnetic-like component in PRF is important. 

\section{Other approaches}

In \CIT{Florkowski:2017ruc} the case of large spin polarization tensor $\omega_{\mu\nu}$ was considered, however, with two additional conditions~\footnote{The conditions \EQn{eq:conditions} were relaxed, for example, in~\CIT{Prokhorov:2018qhq}.}
\beq
\omnL \omnU=2({ \bv \cdot \bv - \ev \cdot \ev }) \geq 0, \quad \omnL \omnUD = -4 \ev \cdot \bv =0.
\label{eq:conditions}
\eeq
In this case one finds
\bel{fpm}
f^\pm_{rs} =  e^{\pm \xi - p \cdot \beta}
\cosh\left(\xi_s\right)\left[\delta_{rs}  - \frac{\tanh\left(\xi_s\right)}{2\xi_s}   \, \bv_\ast \cdot \sigv_{rs} \right]
\eel
where
\bel{zeta}
\xi_s =  \f{1}{2} \sqrt{ \bv \cdot \bv - \ev \cdot \ev }.
\eel
Thus, the quantity $\xi_s$ (multiplied by $T$) can be naturally interpreted as a spin chemical potential, as demonstrated in \CIT{Florkowski:2017ruc}. The applicability of this approach is restricted, however, to particles with momenta satisfying the condition
\beq 
|\tanh\left(\xi_s\right) \frac{\bv_\ast}{2\xi_s} | 
= | \frac{\tanh\left(\xi_s\right) \bv_\ast}{\sqrt{ \bv_\ast \cdot \bv_\ast - \ev_\ast \cdot \ev_\ast }} |
\leq 1,
\label{eq:cond1}
\eeq
where we used $\sqrt{ \bv \cdot \bv - \ev \cdot \ev }=\sqrt{ \bv_\ast \cdot \bv_\ast - \ev_\ast \cdot \ev_\ast }$. The condition (\ref{eq:cond1}) takes a particularly simple form for particles with $|\ev_\ast| \ll |\bv_\ast|$. In this case $\bv_\ast/|\bv_\ast|$ becomes a unit vector showing the direction of mean polarization, while $\tanh(\xi_s)$ defines its magnitude. 

Yet another treatment of spin polarization was introduced in \CIT{PhysRevD.46.4603}, where (using our notation) the following ansatz was made for particles
\beq 
\zeta_+^\mu(x,\pv) = u^\mu n(p)\cdot p - n^\mu(p) u \cdot p.
\label{eq:hakim1}
\eeq 
Here $n(p)$ is a four-vector that is perpendicular to $p$. The form (\ref{eq:hakim1}) does not comply with the requirements discussed above and as such it seems to be quite arbitrary. In particular, it is not clear why the flow vector $u$ appears in (\ref{eq:hakim1}).

\section{Insights from models with classical description of spin}

Different conditions that appear above for the coefficients of the spin polarization tensor $\omega_{\mu\nu}$ and three-momenta of particles $\pv$ indicate that the discussed forms of the equilibrium Wigner functions are limited in their physical applications to some definite range of space-time and momentum variables (let us say in LAB frame). Some light can be shed on this limitation if we refer to a kinetic theory with classical description of spin~\CITn{Florkowski:2018fap}. The classical approach shows that for large spin polarization the systems become anisotropic in momentum space. Such anisotropy has not been addressed yet in present formulations, so this is the work to be done in future studies. Fortunately, the classical description of spin shows also consistency with the forms obtained for small polarization. Consequently, taking together results obtained with the Wigner functions and classical spin description we obtain a convincing physical picture for sufficiently small $\omega_{\mu\nu}$. In the end it is not a very much restrictive constraint, since the measured values of the global spin polarization remain at the level of a fraction of 1\%.

\section{Conclusions}

In this letter we have discussed different concepts and forms of the spin chemical potential entering the formula for the semi-classical equilibrium Wigner function of particles with spin $\onehalf$. Our results suggest using the tensor form of the chemical potential that originates from its use as a Lagrange multiplier in the conservation of angular momentum~\CITn{Florkowski:2017ruc,Becattini:2018duy}. Moreover, recent forms of the equilibrium Wigner function suggest that the spin chemical potential (scaled by temperature) should be small. In this case, the scalar spin chemical potential can be expressed by the tensor form. Interestingly, the scalar form should be momentum dependent, a feature connected with the fact that the spin polarization is always defined in the particle rest frame. 

Several comparison to other works using various concepts of the spin chemical potential have been made. This can help to relate different results and interpretations. Our results can be useful for further development of hydrodynamics of particles with spin $\onehalf$ and serve to interpret experimental measurements of particle spin polarization.

\medskip

{\bf Acknowledgements:} We thank Nora Weickgenannt and Enrico Speranza for many interesting discussions that helped us to identify the problems discussed in this letter. This work was supported in part by the Polish National Science Center Grants   No. 2016/23/B/ST2/00717 and \\ No. 2018/30/E/ST2/00432.

\appendix
\section{Useful formulas and identities}
\label{sec:usefulid}

Our conventions for labels and signs of Dirac bispinors are as follows:
\beq
\usp &=&  \sqrt{E_p+m} 
\left( \begin{array}{cc}   1 & \varphi_s\\ \frac{\sigv \cdot \pv}{E_p+m} &  \varphi_s   \end{array}\right), \\
\vsp &=&  \sqrt{E_p+m}
\left( \begin{array}{cc}   \frac{\sigv \cdot \pv}{E_p+m} & \chi_s\\ 1 &  \chi_s   \end{array}\right),
\eeq
with 
\beq
\varphi_1 =    
\left( \begin{array}{c}   1  \\ 0 \end{array} \right) ,  \quad
\varphi_2 =    
\left(\begin{array}{c}   0  \\ 1 \end{array}\right),\nn\\ \quad
\chi_1 =    
\left(\begin{array}{c}   0  \\ 1 \end{array}\right), \quad
\chi_2 =    
-\left( \begin{array}{c}   1  \\ 0 \end{array}\right).
\eeq
The spin operator $\SmnU$ is defined by the expression
\bel{SDO}
\Sigma^{\mu\nu}  = \f{1}{2} \sigma^{\mu\nu} = \f{i}{4} [\gamma^\mu,\gamma^\nu], 
\eel
which in the Dirac representation gives
\beq
\Sigma^{0i} = \f{i}{2} \left( \begin{array}{cc} 
0 & \sigma^i \\ \sigma^i & 0 
\end{array}\right), \,\,
\Sigma^{ij} = \f{1}{2} \epsilon_{ijk} \left( \begin{array}{cc} \sigma^k & 0\\ 0 & \,\,\,\sigma^k \end{array} \right),
\eeq
with $\sigma^i$ being the $i$th Pauli matrix. The $\gamma_5$ matrix is defined as $\gamma_5 = i \gamma^0 \gamma^1 \gamma^2 \gamma^3$.

Using the above definitions of the Dirac bispinors one can directly derive several useful relations which are listed below. Some of them are well known but the other are rather not popular so we list them all for completeness. With the short-hand notation  
\beq 
X_{rs} = \left( \delta_{rs} + \zetav \cdot \sigv_{rs} \right)
\eeq
one obtains:
\beq
\sum_{r,s}\ubarsp \urp X_{rs} \!\!\!\!&=&\!\!  -\sum_{r,s}\vbarrp \vsp X_{rs} = 4m, \nn \\
\sum_{r,s}\ubarsp \gamma_5 \urp X_{rs} \!\!\!\!&=&\!\! \sum_{r,s}  \vbarrp \gamma_5 \vsp X_{rs} = 0, \nn \\
\sum_{r,s}\ubarsp \gamma^\mu \urp X_{rs} \!\!\!\!&=&\!\! \sum_{r,s}  \vbarrp \gamma^\mu \vsp X_{rs} = 4 p^\mu, \nn \\
\sum_{r,s}\ubarsp \gamma^0 \gamma_5 \urp X_{rs} \!\!\!\!&=&\!\!\!\!\! -\sum_{r,s}  \vbarrp \gamma^0 \gamma_5 \vsp X_{rs} = 4 \pv \cdot \zetav, \nn \\
\sum_{r,s}\ubarsp \gammav \gamma_5 \urp X_{rs} \!\!\!\!&=&\!\!  4 \left(m \zetav +  \frac{\pv \cdot \zetav}{E_p+m}\,\pv \right), \nn \\
\sum_{r,s}  \vbarrp \gammav \vsp X_{rs}  \!\!\!\!&=&\!\!\! - 4 \left(m \zetav +  \frac{\pv \cdot \zetav}{E_p+m}\,\pv \right), \nn \\
\sum_{r,s}\ubarsp \Sigma^{0i} \urp X_{rs} \!\!\!\!&=&\!\! \sum_{r,s}  \vbarrp \Sigma^{0i} \vsp X_{rs} \nn \\
&=& -2 \epsilon^{ijk} p^j \zeta^k, \nn \\
\sum_{r,s}\ubarsp \Sigma^{mn} \urp X_{rs} 
\!\!\!\!&=&\!\!\!\! -2 \epsilon^{mni} 
\left(\!\frac{(\pv \cdot \zetav)\,p^i}{E_p+m}-\!E_p \zeta^i \right), \nn 
\\
\sum_{r,s}  \vbarrp \Sigma^{mn} \vsp X_{rs} \!\!\!\!&=&\!\!\!\! -2 \epsilon^{mni} 
\left(\!\frac{(\pv \cdot \zetav)\,p^i}{E_p+m}-\!E_p \zeta^i \right). \nn \\
\eeq
%

\printcredits

\bibliographystyle{model6-num-names}

\bibliography{cas-refs}





\end{document}